\documentclass{sig-alternate}

\newcommand{\comment}[1]{{\textbf{[\sffamily xxx: #1]}}}
\renewcommand\comment[1]{} 

\usepackage{float}
\usepackage{pifont} 
\usepackage{caption}
\usepackage{subcaption}
\usepackage{url}

\usepackage{algorithmicx}
\usepackage[noend]{algpseudocode}
\usepackage{algorithm}
\usepackage{multirow}

\begin{document}
%

\title{Aashiyana: Design and Evaluation of a Smart
	Demand-Response System for Highly-stressed Grids}

\numberofauthors{2} 
%
\author{
%
%
\alignauthor Zohaib Sharani, Khushboo Qayyum, Affan A. Syed\\
       \affaddr{Systems and Networks Research Lab, National University of Computer \& Emerging Sciences}\\
       \affaddr{Islamabad, Pakistan}\\
       \email{first.last@sysnet.org.pk}  
\alignauthor Noman Bashir\\
       \affaddr{Center for Advance Studies in Energy, National University of Science \& Technology}\\
       \affaddr{Islamabad, Pakistan}\\
       \email{first.last@sysnet.org.pk}
}

\maketitle
\begin{abstract}
This paper targets the unexplored
    problem of demand response within
    the context of power-grids that are
    allowed to regularly enforce blackouts
    as a mean to balance supply with demand:
    \emph{highly-stressed grids}.
Currently these utilities use
    as a cyclic and binary (power/no-power)
    schedule over consumer groups leading
    to significant wastage of capacity and long hours
    of no-power.
We present here a novel building DLC
    system, Aashiyana, that can enforce
    several user-defined
    low-power states.
We evaluate distributed and centralized
    load-shedding schemes  using Aashiyana
    that can, compared to current load-shedding
    strategy,  reduce the number of
    homes with \emph{no} power by $>80$\%
    for minor change in the
    fraction of homes with full-power.
\end{abstract}

%
%

\section{Introduction}


Demand response (DR)
    is a smart-grid technology allowing grid to
    communicate a demand decrease request
    to meet supply,
    against traditional supply-following load behavior,
    using indirect (pricing) or direct (through some control) signals.
Utilities prefer Direct Load Control (DLC) as it
    gives guaranteed reduction but is difficult for user;
    pricing signals leave the customer in charge but have an
    uncertain and possibly time-delayed demand reduction~\cite{Mary09:pricing_confusing}.
Pricing signal can also  result in secondary peaks
    due to behavioral shift~\cite{Rad10:DLCbetter}.

%

We however argue that most DLC work has
    focused on \emph{over-provisioned} grid systems
    of developed countries,
    with a focus on increasing revenue and reliability~\cite{Keshav:Direct_adaptive_control},
    but remains largely blind to
    the unique characteristics of \emph{highly-stressed grids}
    of countries (like Pakistan, Nepal, and India)
     with a very large and nearly continuous supply-demand gap.
As an example, for Pakistan,
    this gap can be as high as 6GW during summers,
    but stays around 1.2GW even during the winter months (2011-2012)~\cite{pak_energycrises_paper}.
The (largely national) utilities
    in these countries enforce
    periodic events of controlled blackouts,
    or load-shedding, to relieve this stress.
Existing DLC mechanism,
    in trying to balance consumer comfort with
    \emph{some} reduction employ fine-grained
    (in both time and amount of load-shed)
    load-control, especially at residential homes~\cite{adam_example:DLC_Summer_Discount_Plan,Peak_Saver}.
Such DLC allows for control events,
    like changing HVAC set-points,
    or possibly for controlling the AC for a few hours
    a day with over-ride facility~\cite{adam_example:DLC_Summer_Discount_Plan}.
While these mechanism are quite useful
    in shaving off consumption peaks and
    preventing peaker plants
    from running (thus saving money),
    they are \textbf{inadequate in their magnitude as well
    as flexibility}
    for managing the large and continuous
    gaps that exist in
    highly-stressed grids.
\comment{Noman, can we get SOME cite for the amount of stress
on our grids? with a citation! -af}

We believe that the consumers in a highly-stressed grid
     --- being acclimatized to frequent blackouts ---
    are much more amenable to  aggressive
    DLC mechanisms and thus willing to accept
    a wider-range of load-shedding policies.
This demand reduction,
    however,
    will have to be done through some automated
    system as users cannot be expected to
    manually respond to any, potentially large,
    load reduction signal.
\emph{We thus propose instrumenting homes with a system
    that provides utilities with transitions
    to several \textbf{low-power states}
    that map to user-specified appliances.}

%
%
%


In this paper we design
    and evaluate a novel and practical
    home-level DLC system
    solution, \emph{Aashiyana},
    that can implement several
    user-configurable power-states
    of a home.
This system is practical
    as it can retrofit into
    the existing wiring scheme of homes;
    is of low cost while controlling most
    appliances in a home;
    provides home consumers a flexible
    way to describe these
    lower-power states as a compact
    disconnectivity matrix
    requiring one-time configuration.

A question remains
    regarding incentives for  power utilities
    to promote a proliferation
    of the Aashiyana DLC system when their
    current strategy of full blackouts is working?
We believe that for national utilities
    with huge demand-supply gap,
    possibility of social unrest and potential political
    backlash (for example, road-blockades and tire burnings~\cite{Protest_loadshedding_dawn})
    provides an impetus for government
    to explore alternate solutions.
Aashiyana's penetration enables
    flexible and fine-grained load-shedding policies
    that will reduce the underload wastage from
    the current strategy of coarse-grained, group-level
    shutoff while increasing
    social comfort within
    the \emph{same supply-side constraints};
    a push for such schemes
    will thus come top-down for socio-political reasons.
A bottom-up push will come
    as consumer penetration of
    Aashiyana homes increases,
    and people observe the increased
    comfort level of their neighbors.

A serendipitious benefit of our DLC mechanism
    would be to actually reduce the load on the grid
    by removing the need for battery backups,
    extensively used already in countries with a stressed grid.
These backup solutions use
    inefficient battery storage to
    transition into a single
    ``low-power'' state,
    but have shown to exacerbate
    the supply-demand gap that  leads
    to greater penetration of battery backup
     and even greater stress ---
     a death spiral for these grids~\cite{Hidden_Cost_Seetharam,SOFTUPS_Sharani}.
Our power-control system will provide
    an exact substitute for these backups,
    but with no inefficiency since
    it gets its (lower) power \textbf{directly from the grid}.

The contribution of this paper
    are the following.
We present the design and implementation
    of \emph{Aashiyana}:
    a novel home-level DLC system that
    can retrofit into wiring system
    to enforce different power-states
    for a home, with the set of appliances
    allowed in each state defined by individual
    home-owner (Section~\ref{sec:Ashiyana}).
This provides a first,
    to the best of our knowledge,
    practical DLC mechanism
    for the dynamics of load-shedding
    of  a highly-stressed grid.
We propose
    two different DLC strategies
    that can be implemented now over a smart-grid
     and leverage any penetration level of Aashiyana
     to improve the social utility without requiring
     increased supply (Section~\ref{sec:algos})
Finally,
    we build a custom simulator to model
    a stressed grid and  evaluate our
    DLC algorithms using the flexibility offered by Aashiyana
    to show that for 90\% penetration
     we can decrease by >80\%
     the fraction of homes with \emph{no power},
     without significantly decreasing
     (in some case actually increasing)
     the fraction of homes with full-power (Section~\ref{sec:Eval_Results}).


\section{Background to Load-shedding}
\label{sec:load_shed_model}

In this section we  provide an overview to the
power grid architecture and players in the power system
in Pakistan, and how the demand-supply gap is monitored
 and converted in load-shedding schedules.
While the exact details might vary across
    different countries, this description
    will give the overall flavor of the problem.
%
%

\subsection{Electrical Power System in Pakistan}

In Pakistan, power sector is primarily state owned,
    with some private sector responsible for generation.
These generation units (GENCOs) work
    with the National Power Construction
Corporation (NPCC), a central body which monitors the
power grid, to asses the overall demand and operate
power plants accordingly.
NPCC, based on a survey done during a time with no supply shortage,
    decides on the allocated split of power generation
    across the ten major (state owned)
    distribution companies (DISCOs).
%



\subsection{Current Load-shedding Schemes}

Pakistan has, due to several socio-political
 reason, become a country which faces a nearly year round
    supply-demand imbalance situation~\cite{pak_energycrises_paper}.
Whenever demand exceeds supply,
    a large segment of consumers (forming a group/zone)
    is shutdown to reduce the overall load on the grid.
The task of enforcing blackouts is accomplished at
two levels: by the DISCOs and in the extreme case the NPCC.

NPCC has an estimate of generation
    based on their dispatch request for the upcoming
    24 hrs.
According to that  estimate, each DISCO is
 allocated a certain budget (based on a ratio previously determined)
 and asked to not exceed that threshold,
 whatever might be the actual demand.

DISCOs also have an estimate of consumption
    from their allocated area.
If this estimate is greater than their share,
    they initiate a schedule of load shedding
    across collection of feeders clustered into
    a zones or groups.
They implement a time-disjoint blackout
    schedule across these groups, and increase
    the number of hours until they meet their quota.
It is fine if DISCOs to under/over-use
    slightly beyond their quota,
    as long as the over all grid remains stable.
However,
    if the grid is becoming unstable
    due to overuse,
    the NPCC has the nuclear option of shutting down
    the 132kV line to the offending DISCO(s) and restore
    balance to the grid.
In the rest of the paper,
    to simplify the analysis,
    we consider the case of a DISCO
    with a particular quota as representative
    of this more complex load-shedding strategy.

\section{Related Work}

The vast majority of research
    in demand response  has considered
    the issue to shifting peak demand
    to allow for flattening such peaks.
There is, to the best of our knowledge,
    no work that evaluates how such algorithms
    will work when applied to a grid
    where there is a continuous demand-supply gap.
Our work proposes
    a fine-granularity DLC mechanism
    that is practical and leverages the conditioning
    of consumers to  to full blackouts in countries
    with highly-stressed grids.
We divide our related work which focus on
     two different areas:
    grid-assisted DLC mechanisms and
        home-consumption changing systems.

Utilities have, for a long period of time,
    experimented with DLC mechanisms to share peak loads~\cite{adam_example:DLC_Summer_Discount_Plan,Peak_Saver}.
These systems in the US and Canada, respectively,
    give consumers rebates for installing equipment
    to control a specific high power device (like A/C or heating units),
    for few hours a day.
This granularity of control and load-reduction,
    even with high penetration,
    can not meet a \emph{sustained}
    supply-demand gap in \emph{giga-watts}
    that is typical for the highly
    stressed grids of countries our work targets.
Perhaps the most similar work to
    ours is~\cite{Keshav:Direct_adaptive_control,Chandan:2014:ICG:2602044.2602062}.
Keshav and Rosenberg~\cite{Keshav:Direct_adaptive_control}
    propose a smart-grid where
    consumers contend to turn on appliances (proactive)
    or close the last appliance turned on that induces instability (reactive).
iDR~\cite{Chandan:2014:ICG:2602044.2602062}
     proposes a theoretical framework that generates
     a signal for DR with the appropriate amount of
     reduction required from a consumer, such that
     the overall utility is maximized.
Both these work do not have a practical
    system designed to implement their DR mechanisms,
    and specifically do not consider the large and continuous demand-supply gap
    of highly stressed grids.

The second area of research
    focuses on designing appliance or home
    level power reduction --- while considering
    the comfort and ease of home owners~\cite{JPLUG_Bapat,ganu:nplug:,Peak_Demand_Domestic_Appliances:2012, smartcap:2012}.
Both Yupik~\cite{JPLUG_Bapat} and n-Plug~\cite{ganu:nplug:}
    propose adding smart plugs to a few (deferable)
    appliances, whose usages patterns are monitored
    to present appropriate slack when a grid-stress event (demand peak)
    is indicated, using prices in Yupik and frequency in n-Plug.
Smartcap~\cite{smartcap:2012}  uses programmable
    switches or smart-appliances to control
    background appliances using a least-slack-first algorithm
    to flatten any peak.
Srikantha et al.~\cite{Peak_Demand_Domestic_Appliances:2012}
    evaluate how peaks can be flattened if the elastic component
    is allowed to be programmatically controlled.
All these work seek to flatten load such
    that user prefernces are minimally affected;
    they fail to leverage (as they do not consider the problem domain)
    the experience of consumers in highly stressed grid that consequently
    face full blackouts for $>$ 12hrs a day~\cite{pak_energycrises_paper}.
Furthermore, 
    these are all primarily peak-shifting algorithms;
    for a highly-stressed grid there is \emph{a permanent peak}.

To summarize,
    to the best of our knowledge,
    no work has yet considered a) the practical
    concerns of cost and usability for large scale DLC in homes
    and b) leveraged the potential readiness of consumers
    in highly stressed grids for wildly different modes
    of demand management i.e. several levels of fixed power budgets.

\section{Goals and Challenges for a Practical Home-level DLC Solution }
\label{sec:Goals}

We propose to explore flexible and fine-grained
    load-shedding strategies by enforcing
    multiple power states inside a home.
To achieve such control and enable a common house
to implement multiple power states, we needed to
add into the \emph{existing}
 electrical infrastructure.
However, for  practicality
    and usability concerns,
    we define three important goals
    that we want such a power-control system to achieve.
To achieve this, we identified three important
goals for implementation of Aashiyana.

\begin{description}
  \item[Low Deployment Cost (G1)]
We want to control the power state of homes
    using a system that costs around (\$250-\$300).
This price range corresponds to the
    cost of a battery backup,
    which provides a very
    low-power state from stored energy,
    that people are already comfortable purchasing.
This is a challenge since making
    individual devices intelligent
    (using intelligent plugs or power strips ~\cite{ganu:nplug:,JPLUG_Bapat,ZwaveSMARTSTRIP})
    is quite an expensive proposition.

\item[User-friendly Configurations (G2)]
To ensure an acceptance of a power-control
  system, the ease with which a user
  can configure the system is essential.
  This can also correspond to a visual
  representation of the benefit for performing
  these configurations.

  \item[Multiple Communication Strategies (G3)]
In order to handle different mechanisms
    for load-shedding, we want our
    designed system to support several communication
    capabilities.
Thus we would like homes within a neighborhood
    to communicate and also a grid-scale
    rendezvous mechanism for a centralized algorithm.
\end{description}


We next use these goals
    in guiding the design of Aashiyana:
    a prototype power control system that can retrofit\footnote{An abundance of smart-appliances is not
    a sound assumption in the target countries of interest.}
    into the wiring structure of a home,
    and provide utility companies with an ability
    to reduce demand to guaranteed levels
    while providing consumers a way  to easily configure
    their homes for each power level.

\begin{figure}[htp]

\includegraphics[width=0.47\textwidth]{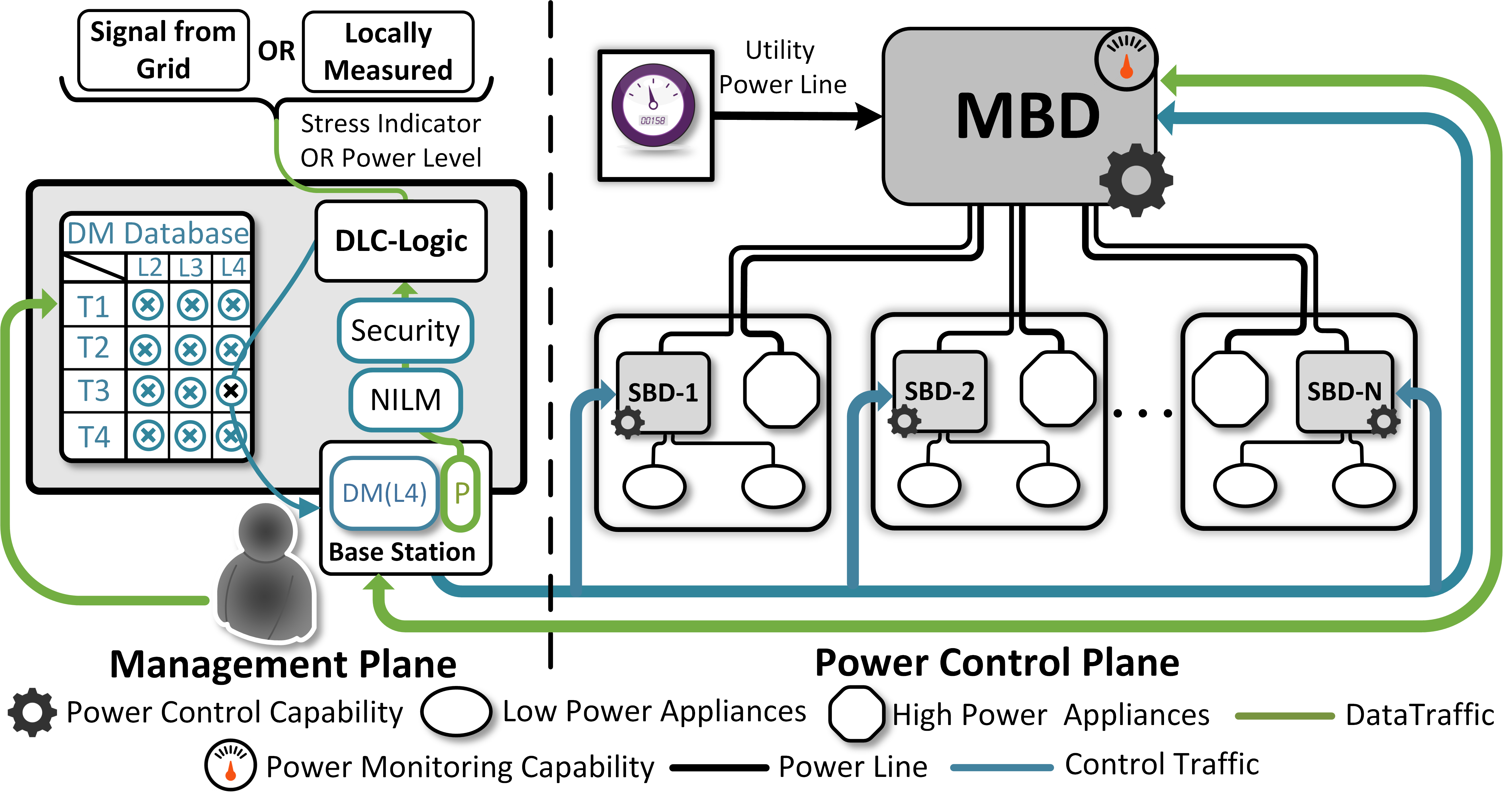}
\caption{Aashiyana Architecture}
\label{fig:Aashiyana}

\end{figure}

\section{Aashiyana: A practical system to implement power control }
\label{sec:Ashiyana}

Aashiyana is meant to be a user configurable
    system to enforce demand response that
    is acceptable to both utilities as well
    as consumers.
Our major focus is,
    inline with the goals above,
    to design and system that can enforce
    a consumption budget at each home,
    while allowing the users the ability
    to flexibly configure devices running
    at each demand reduction level.
We first describe the major design decisions
    for Ashiyana and then the
    architectural components of our
     demand-management solution.

\subsection{Design Decisions for Aashiyana}
\label{sec:design_decisions}
A first design decision
    was to select the location
    and granularity of appliance control
    within a home.
The overarching constraint in this decision
    was the cost constraint.
For granularity we decided to restrict
    control to the level of switching
    devices on-or-off.
Instrumenting any greater level of
    smartness would require significantly
    larger cost and configuration complexity.

We decide on locating our
    two control components at the
    main distribution board and
    at the level of switch boards
    installed at each home,
    in light of the traditional wiring
    structure for Pakistan (Figure~\ref{fig:Home_Wiring_Scheme}).
These locations provides us sufficient
    control as the high-power device
    sockets (separate in each room for A/C, Refrigerators)
    are accessible from the main distribution board,
    while individual sockets as well as fixed appliances
    like fan, and lights for each are accessible
    from a switch box.
The choice of these location,
    and our custom board design,
    leads to a price point of around \$300 for
    a four room home (details in Section~\ref{sec:Ashiyana_impl}).


We next decided to
    restrict the power consumption of
    our homes to \textbf{five levels}.
\emph{Level 5 and Level 1} represent
    the current two modes of
    unrestricted\footnote{limited to meter rating}
    power consumption and full disconnection, respectively.
\emph{Level 4-2} represent power consumption
    that is 75\%, 50\%, and 25\%
    of full rated capacity.
We restrict ourselves to just three configurable
    level to ensure ease-of-use (G2)
    as a user will have to supply,
    for each level and possibly even different
  time-of-day or week,
    a matrix of control points that will be disconnected
  (which we call a \emph{Disconnectivity Matrix (DM)}).

Finally,
    we also decide on using existing building automation
    and IoT frameworks, like ~\cite{Dixon:HomeOS,OpenBAS}
    to enable a ease of application development and
    a robust rendezvous mechanism.
While it is conceivable to have built
    a custom system to provide redirection
    and web service accessibility,
    the development overhead and reliability concerns
    tilted us in favor of our final decision.

To better explain the architecture of Aashiyana,
    we split it into two planes (Figure~\ref{fig:Aashiyana}):
    Management Plane and the Power Control Plane.
While the home management plane implements
    the logic to trigger different power-levels,
    the power control plane enforces these states.
We assume an external grid management plane
    that is (optionally) responsible for providing
    the demand-reduction signal to homes
    that implement Aashiyana.
We detail the components
    in these planes next.

\subsection{Power Control Plane}
The power control plane consists
    of components that enable the enforcement
    of different power-states of a home.
As we already argue in previous section,
    we have decided to control the on/off
    state of appliances at the level
    of switch boards and the main distribution board.
For this purpose,
    we need two different control
    components:
\begin{description}
  \item[Main Board Device (MBD)] This component
  is primarily responsible for controlling
  all heavy appliances from one location.
  This control is possible from the main distribution
  box where, as per current wiring strategy,
  each high power sockets is connected via
  separate (higher rating) wires and circuit breakers
  (c.f. Figure~\ref{fig:Home_Wiring_Scheme}).
  The MBD can similarly control power supply
  to every room as a single wire goes to the
  switch box of each room, and distributed
  to individual appliances from there.
  A final purpose of the MBD is to monitor
  the power consumption at each room level
  to provide monitoring ability to prevent
  overuse  at room level. We can increase the observation
  granularity to socket or appliance level by
  using NILM techniques in the management plane,
  aided  by state info about switches in each room. We do not use per switch
  power monitoring to keep the system-cost down (G1).

\item[Switch Board Device (SBD)] This component
 is located inside the switch board for each room,
 which terminates the direct line coming from the
 main distribution box. It is responsible,
 much like MBD, to control the wires distributing
 from this sockets. As shown in Figure~\ref{fig:Home_Wiring_Scheme},
 some wires go to hard-wired
 devices like fans and lights, while the rest go
 to individual sockets. We thus will have to \emph{infer}
 devices connected to the sockets, while
 the hard-wired devices can be one-time configured.
\end{description}

Both these components communicate their
    data (power, state) to the home management
    plane through some IoT-based communication
    technology. We describe this plane next.

\subsection{Management Plane}

The Management Plane represents
    the brains behind the power-management
    of Aashiyana.
This plane consists of a DLC-logic
    module as well as Base-station
    component that enables the communication
    using the IoT technology (802.15.4,
    Z-wave, power-line) used by the MBD and SBD.

This plane is first responsible
    saving user preferences
    in the form of a database of
    disconnectivity matrix (DM) for
    each power-state.
It collects consumption data
    from the control plane and provides
    this information to users for easy
    selection (in line with G2) of
    a DM that meets a given power budget.

A second, and most important,
    function is to appropriately
    respond to a grid-stress signal.
Thus, any DLC algorithm is implemented
    through the selection of the appropriate
    power-state by the DLC-logic present
    within this plane.
Once the power level is selected the appropriate
    DM is used to send commands to the control
    plane in order to switch off power to selected
    points.

A final responsibility
    of the management plane is to perform
    cross device power monitoring.
This capability is needed,
    as discussed above,
    to perform NILM across the SBDs.
Similarly,
    tamper-detection and prevention strategies
    will also be implemented at this central
    location.

\subsection{Grid Management Plane (optional)}

The demand-reduction process initiated
    by the management plane requires an indication of grid-stress.
While this detection can occur
    in a fully distributed manner at each home
    (by, for example, sensing frequency~\cite{ganu:nplug:}),
    we expect the utilities to have some
    demand-response logic based on
    the current supply-demand gap.
This logic is represented by the
 optional Grid Management Plane (not shown in Figure~\ref{fig:Aashiyana}).

Using the Aashiyana system,
    the utilities now have the option
    to specifying five power-states
    (as described in Section~\ref{sec:design_decisions}),
    instead of the current two states
    of uninhibited consumption or a full blackout.
We reiterate that these different power states
    are  acceptable for
    consumers where black-outs
    are a regular occurrence (highly-stressed grids),
    but might not be equally palatable for
    consumers where such events are unthinkable.

\section{Aashiyana: Implementation and Evaluation   }
\label{sec:Ashiyana_impl}


\begin{figure}[tp]
    \centering
    \includegraphics[width=0.45\textwidth]{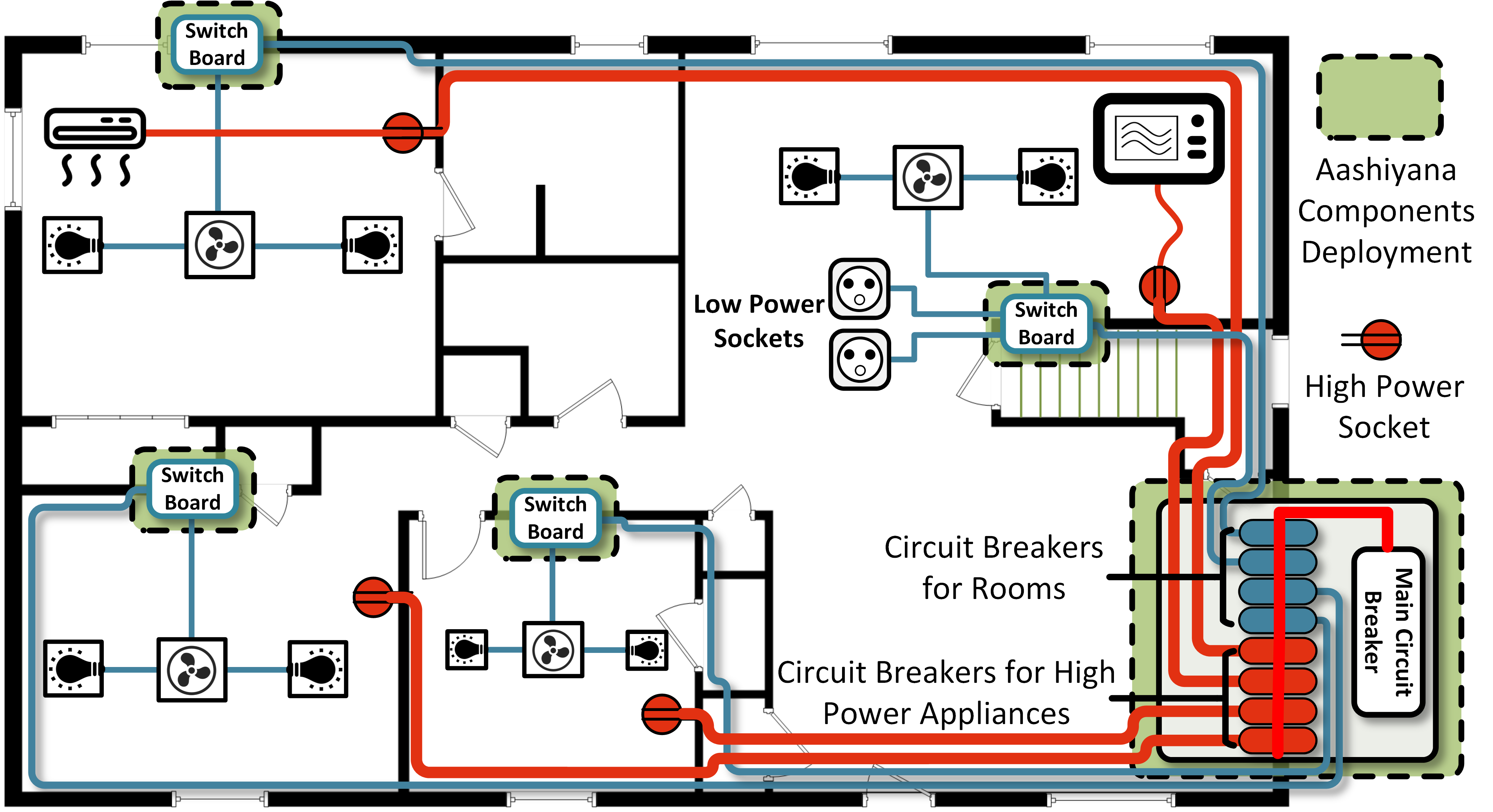}
    \caption{Home Wiring Scheme}
    \label{fig:Home_Wiring_Scheme}
\end{figure}

In  this section we present our working
    prototype of an Aashiyana system.
We first discuss
    the lab-prototype home we built
    to emulate the wiring scheme of a home,
    followed by the implementation details
    of the hardware devices, as well as the
    software and communication stack to implement
    out system.
We end the section with evaluation
    of the latency, reliability,
    and power consumption for our system.

\subsection{Prototype Home}

\begin{figure} [htp]
\centering
   \includegraphics[width=0.4\textwidth]{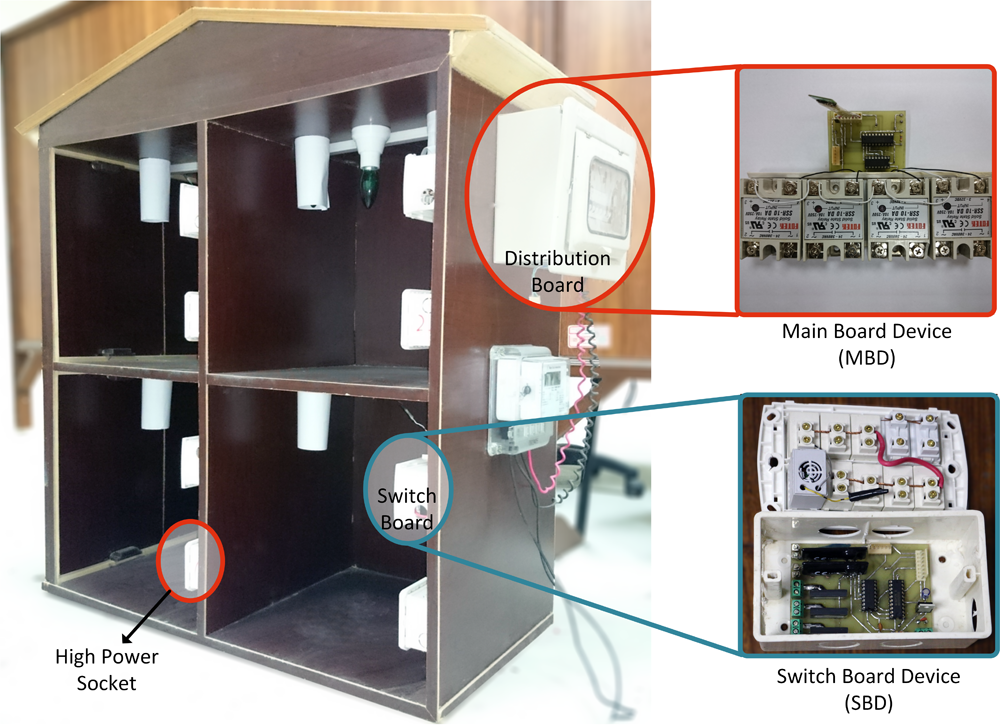}
    \caption{Aashiyana Prototype Home}
    \label{fig:Aashiyana Prototype Home}
\end{figure}

Figure~\ref{fig:Home_Wiring_Scheme}
    shows the wiring of a typical home with two
    types of wires going to each room from the
    main distribution box.
The first type are the high-rating wires that
    go to special sockets used to connect high
    power devices in each room.
The second type of wires terminates in each room
    at the switchboard level and spreads around to
    fixtures (fans and lights) as well as sockets.
We replicate this wiring, by enlisting a professional
    electrician, in a prototype and scaled-down
    four room home (Figure~\ref{fig:Aashiyana Prototype Home}).
This home has a main distribution board with
    circuit breakers for each room and high power appliances.
A smart-meter, courtesy of microtech~\cite{MicroTech_website},
    is installed to observe energy consumption at the home level.
However, due to a proprietary protocol,
    we are currently using the current-cost power meter
    to obtain this consumption data.

\subsection{Hardware Devices}

The \textbf{Main Board Device (MBD)} located inside the distribution
box of the home consists of three major components; a communication
module, a processing unit and controlling relays. We had wi-fi,
PLC, and RF as candidates for communication technologies. Keeping
in mind our cost constraints, we opted for a CC2500 RF module,
having cost of \$3.95, for this purpose.
We use  msp430 launchpad modules(\$2.80 each)
     for our processing needs.
We chose Solid State Relay (SSR)
    over Electromagnetic Relay (EMR)
    due to its longer life and noiseless operation
    despite its slightly higher cost.
Four high rating (25A @ 220V, \$7.94 each) SSRs
are used to control the high power sockets.
We also design our PCB board to ensure that the MBD
    fit within the space
    constraints of the main distribution board.

The \textbf{Switch Board Device (SBD)} is similar
    to the MBD with a primary
    difference in number of relays and their ratings.
The SBD has five outlets with
    three of them for fixed appliances (fans and lights) and two power sockets.
We used 1A rating SSR (\$5.18) for the three (known rating) appliances
    and 5A ones (\$14.56) for power sockets with variable load.
We assume room level control by shutting off
    all relays of an SBD.

The \textbf{Base Station} is responsible for implementing
    the control-decisions of the management plane
    and send them to the MBD and SBDs
    using the communication protocol over the CC2500 RF chip.
This chip is coupled with a micro-controller,
    currently an mbed, but is being
    replaced with our msp430 option for cost savings.

\begin{figure}[tp]
      \centering
      \includegraphics[width=0.3\textwidth]{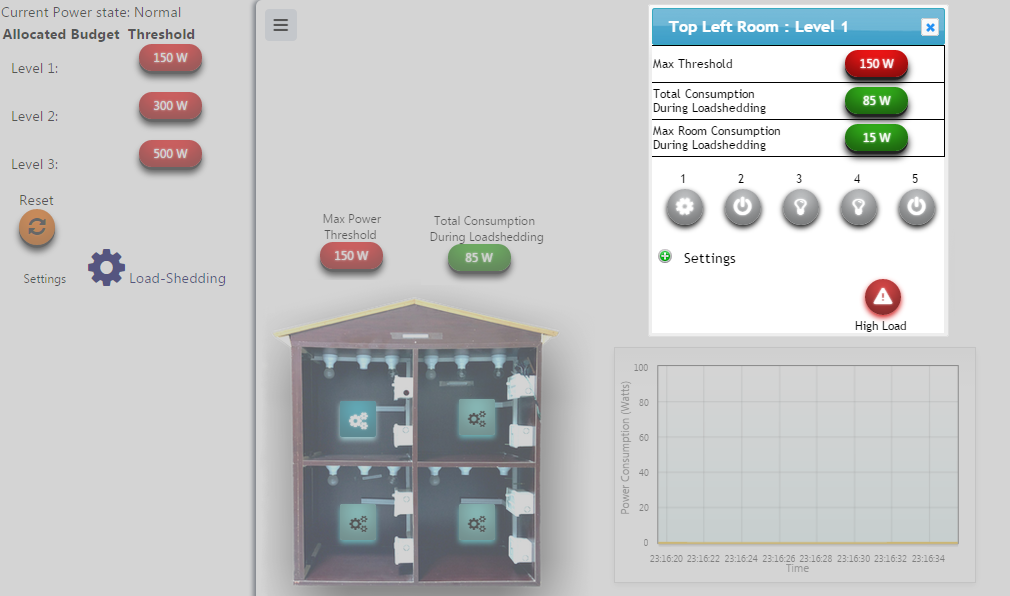}
      \caption{Aashiyana Application}
      \label{fig:Aashiyana_Application}
  \end{figure}
\subsection{Software Components}

We decided to use Microsoft Research's
    Lab-of-Things (LoT), an  open source platform
    that enables seamless integration of home-automation systems
    into a coherent framework by requiring vendor specific drivers.
It further provides a cloud-based rendezvous
 mechanism and a relayering feature,
 allowing seamless NAT traversal capability so
 that applications built are accessible from anywhere.
We wrote a LoT application for Aashiyana system which
 implements a the disconnectivity matrix
 when it receives the external stress signal.
User interface (UI) of the application (Figure~\ref{fig:Aashiyana_Application})
 is developed as an HTML5 web-application, and is useful to input
 the disconnectivity matrix as well as view status
 from any internet connected device.
Currently, we use a dedicated
 laptop as a home hub but target a
 low-cost tablet supporting .NET
  framework for around \$65 \cite{tablet_ref:Low_Cost_Tablet}.



To integrate our hardware system to the LoT framework,
    we custom wrote the device driver and scout for our
 Aashiyana system.
Device scout is
 responsible for discovery of Aashiyana control devices.
The device drivers then allow the application to view
    each control point as a role that can be mapped to a
    particular disconnectivity matrix.
These
  device driver talk with the base station through a serial port,
  and the base station uses its communication
  stack to coordinate with the SBDs and MBD.
Our current implementation thus,
    besides implementing different power states,
    also provides LoT roles that allow
    users to switch on or off specific devices
    as an additional feature.


The communication stack has a discovery part (used by the LoT scout code)
    in which the base station scans for all home-id specific
    devices in the vicinity at bootup
    and stores their addresses in a table.
Each devices is assumed to have a
    unique ID assigned (starting from 1)
    and a disconnectivity matrix is converted
    into a command to switch off corresponding
    relays at each control device.
This command is sent as a broadcast message,
    where the relay state is represented by bits in a byte.
We thus use just first five bits (corresponding to 5 relays)
    of a byte.
Each device extracts the information byte
    corresponding to its device ID and implements
    the state described by the binary values in that byte.

We require  an acknowledgment from every
    device for reliability.
The ack contains the current state of the relays
    on that device allowing the LoT application
    to have an accurate view of the home-power state.
The base-station
    retries three times if ack from any device is not received.
If the receiver
    fails to respond altogether,
    the base station send a nack status to the application
    to potentially show an error message.
We prevent collision between the acknowledgment by implementing
    time-disjoint slots of 5ms based on the unique device IDs.
The base station timeout is 21ms to cater for
    the five devices in our prototype home.

\subsection{Practical Evaluation of Aashiyana}

We now evaluate the practicality of our system
    by measuring three important and practical
    parameters: power-level switching  delay,
    communication reliability, and power consumption.

\subsubsection{Switching Delay}

Since the purpose of Aashiyana is
    to have a system that can help
    attain balance with the demand-supply
    gap, its reaction time has to be on
    the order of reaction delays for a grid stability.
A power grid requires an instantaneous
    matching of supply with demand:
    regulating reserves\footnote{Regulating Reserve is the capacity of generators to supply energy within an economic dispatch
    interval in response to the grid frequency variations.},
    however, are typically employed to manage the time
    until load-following reserves can come online.
A typical grid has around \emph{five minutes}
    of this \emph{stability time}, but can also reach up to an
    hour in some cases~\cite{NREL_OperatingReserve}.

We thus would like to ensure that our system responds
    to stress signals at least an order-of-magnitude
    quicker than this stability time to allow allow grid-balancing
    using any iterative demand response algorithm
    (with possibly multiple round-trip delays).
In order to compute this delay, we measure
    the time from selection of a particular
    Disconnectivity Matrix (DM)
    to when the corresponding state is implemented.
We measure this delay in two parts, using
    different experiments.

First, we observe the latency introduced by the
    Lab-of-Things framework in processing this request.
We randomly generate DM at a delay of around 1 minute,
    use the LoT logger module to time-stamp (at msec granularity)
    the entry into our implementation code,
    as well as the exit from the driver for our base-station
    (just before writing to the base-station serial port).
Over one thousand samples,
    we observe typical delay of around 1-2 msec,
    but occasionally (due to OS context switching)
    a worst case delay between 5-8 msec.

The second part of delay
    is measured from the instant the base-station
    receives the DM until a control device (SBD or MBD)
    switches off a relay.
To measure this delay,
    we time-stamp on the base station
    the time when we receive the first byte
    of control data (including the DM)
    from the serial port (operating at 9600 baud rate),
    to the time when the target device switches off the relay.
This switching event is detected by connecting
    the GPIO output controlling the relay back to an extra port
    on base-station.
Notice that we ignore the acknowledgement time
    from this measurement since as soon
    as the relays have switched the load on
    grid has been appropriately reduced.
However, for cases where
    the control packet is lost,
    timeouts are added into the
    average response time.
We find the average the delay over 120 measurements,
    when sending a control packet containing the DM,
    to be around 28 msec.

Thus the average delay for our system (shown in Table~\ref{tab:Aashiyana_parameters})
    to respond to a stress signal is around 30 msec,
    if we consider internet-network latency of around 200msec;
    our response time is more than \emph{two-orders-of magnitude} faster
    that the grid stability time.

%

\subsubsection{Communication Reliability}
%

We now assess the reliability of the communication
    between the base station and switching devices,
    since loss of control packets will impact the over all
    grid stability.
This reliability is a function of communication range,
    hence we assess the packet reception rate (PRR)
    for our system over different range.
Table~\ref{tab:Aashiyana_parameters} shows  94\% PRR,
    over 100 iterations, at the range of 50m.
While not currently implemented,
    we believe the a mesh network that provides
    99\% reliability (like CTP~\cite{ctp:Sensys2009})
    is best suited to provide guarantees over such lossy links.

\subsubsection{Power Consumption}

Finally,
    we observe the power consumption
    of our prototype system.
We need this value to be a nominal amount
    so as to best manage the limited budget assigned
    to a home.
We measure that an SBD consumes 0.4W,
    with all five appliance on,
    but consumes 0.1W with devices off,
    during a load-shedding event.
An MBD, with higher rating relays,
    consumes 0.36W and 0.1W
    during these two states.
Thus for a four room home,
    the total power cost in implementing
    L5 is 1.96W, while L0 requires 0.5W.

%

\begin{table}[tp]
\resizebox{0.47\textwidth}{!}{%
\begin{tabular}{c|ccl}
\multirow{3}{*}{\textbf{Latency}}       & \textit{\begin{tabular}[c]{@{}c@{}}Software \end{tabular}}       & 5ms                &                                        \\
                                        & \textit{\begin{tabular}[c]{@{}c@{}}Hardware \end{tabular}} & 28ms               &                                        \\
                                        & \textit{Total}                                                          & 33ms               &                                        \\ \hline
\multirow{4}{*}{\textbf{Reliability}}   & \textit{Distance (m)}                                                   & \textit{PRR}       &                                        \\
                                        & 10                                                                      & 100                &                                        \\
                                        & 25                                                                      & 98                 &                                        \\
                                        & 50                                                                      & 50                 &                                        \\ \hline
\multirow{5}{*}{\textbf{Power \& Cost}} & \multicolumn{1}{l}{}                                                    & \textit{Power (W)} & \multicolumn{1}{c}{\textit{Cost (\$)}} \\
                                        & \textit{MBD}                                                            & 0.36                & 40                                     \\
                                        & \textit{SBD}                                                            & 0.40                & 53                                    \\
                                        & \textit{Home Hub}                                                       & 3                   & 65                            \\
                                        &\textit{Total (4 room)}                                                           &
                                        5                   & 317

\end{tabular}
}
\caption{Aashiyana Home Evaluation Parameters }
\label{tab:Aashiyana_parameters}
\end{table}

 With these values, the power of our Aashiyana prototype, if all devices are operating, will be an additional 2 watts and much less when a lower power-level is enforced.

%
%
%
%
%
%

\section{Utility and DLC algorithms}
\label{sec:algos}
We now lay out new load-shedding
    strategies that are made
    possible with an Aashiyana home
    that can operate at different power-levels.
We note that currently
    load shedding is implemented at group level
    where all homes under the feeder
    belonging to a group are set to L1 (no power).
Since Aashiyana system will be
    introduced incrementally,
    we consider different
    load-shedding (aka DLC) policies under
    varying \emph{Aashiyana Penetration (AP)} levels.

In order to evaluate the benefit
    of different strategies we first
    model the social \emph{utility}
    for a home to operate at different power-levels.
We then propose two different
    load-shedding algorithms to manage
    the varying demand supply gap.

\subsection{Utility Function for Different Power-levels}
\label{sec:Utility_func}

We base our utility function
    definition on two observations.
First,
    for a consumer whose home is
    shifted to \emph{any} restricted
    power state (L4-L1),
    the loss of utility is large.
The second,
    and mirror observation,
    is that for a consumer who is
    currently accustomed to having complete shut-off
    \emph{any} power to run basic appliances
    is appreciated.
Representing these
    state changes with
    two specific thresholds $Th_U$ and $TH_L$, respectively,
    and considering a linear utility variation between
    them we define our utility function as:

\[U(U_{max},th_U,th_L)=\left\{\begin{matrix} U_{max} & for~L5\\ th_U & for~L4 \\ \frac{th_U+th_L}{2}& for~L3\\ th_L & for~L2\\ 0 & for~L1 \end{matrix}\right.\]

Different values of threshold correspond
    to different utility interpretation that we will evaluate
    in the next section.

\subsection{Load-Shedding Algorithms}

Given the above utility function,
    we believe that DLC algorithms
    for load-shedding can provide
    much greater utility as opposed
    to the classical option of completely
    shutting off power.
We propose two different algorithms,
    one central and the other distributed
    that try to maximize the utility
    offered to the customers.
Our algorithms enforce reduction
    at an hour-long granularity,
    and we ensure (unless impossible)
    that customer are not successively
    put into a \emph{any} load-shedding state.
Similarly,
    we limit Aashiyana homes to L2,
    leaving L1 (full shutoff) as a final resort,
    using an emergency signal.
We also only use additive backoff;
    any increase (multiplicative or additive)
    to meet extra supply made available due to reduction
    were deemed too prone to oscillations.
With a grid trying to \emph{meet} its demand,
    it thus appears that only decrease in demand,
    returning to full level at the end of an hour,
    is the appropriate response.

It is important to note here that
    utility companies,
    predominantly state owned in countries
    requiring load-shedding,
    will be motivated to decrease the discomfort
    (and thus potential for unrest)
    by promoting the deployment of
    Aashiyana-like systems.
Consumers,
    however, will be motivated when
    they see their neighbors with the
    system installed having greater utility.
While they can buy
    and install local backups (that have
    fixed and recurring cost),
    our system can be deployed
    at a similar cost
    but fewer power units consumed
    (no losses of a battery backup),
    for the same effective utility.

\subsubsection{Distributed DLC Algorithm}
\begin{algorithm} [tp]
 \caption{Distributed DLC algorithm}
 \label{alg:DDA}
\begin{algorithmic}[1]
\Require $sl,DP$
\State $\forall h_i \in \mathbb H_a $

   \If{$(LS_{lh} == true \> AND \> Emergency == false)$}
     \State$exit$
   \EndIf
\State$ r.h_i \gets rand(1,100)$
   \If{$(DLCDone == false)$}
     \If{sl<5}
     \State$sl \gets 5$
     \EndIf
    \State$h_i.sl \gets sl , h_i.sl_{init} \gets sl $
        \If{$(r.h_i<h_i.sl) $}
        \State $DLCDone == true$
             \If{$(r.h_i>(1-DP.\alpha_{L4})h_i.sl)$}
             \State$cl \gets L4$
             \ElsIf{$((DP.\alpha_{L2})h_i.sl <r.h_i<(DP.\alpha_{L3}+DP.\alpha_{L2})h_i.sl)$}
             \State$cl \gets L3$
             \Else
             \State $cl\gets L2$
             \EndIf
        \EndIf
  \Else{$(h_i.sl \gets h_i.sl_{init})$}
        \If{$(((r.h_i<h_i.sl)\>OR\> Emergencey)\> AND \>  cl \neq L2 )$}
         \State $cl\gets cl-1$
        \EndIf
  \State at the end of hours $DLCDone == false$
  \EndIf
\end{algorithmic}
\end{algorithm}

In our distributed algorithm,
    the power throttling level is stochastically
    generated inside the DLC logic.
However,
    due to a variable Aashiyana Penetration (AP),
    the grid utility is still involved to fully shut-off
    non-Aashiyana homes.
The only information external
    information required is the
    stress level on the grid,
    communicated either by the utility
    or by locally sensing supply frequency~\cite{ganu:nplug:}.
We define this \emph{stress level}
    as $sl= \frac{D-S}{D}$;
    where $S$ is the current supply
    while $D$ is the actual (unfulfilled) demand.
Intuitively this is the fraction of
    actual demand not currently being met,
    and thus the fractional amount to be shed.
%
We propose an iterative algorithm,
    where in the first iteration  an Aashiyana homes
    run a stochastic algorithm (Algorithm~\ref{alg:DDA})
    to determine their new power state.
This state is selected as a function
    of the $sl$ and a
    \emph{Distribution Profile (DP)}
    characterizing the relative percentage
    of homes that we would want to stay
    in L4, L3, and L2.
If, after this load reduction,
    at the start of second iteration\footnote{each iteration is set to be 1 sec,
    sufficient for an Aashiyana response and grid stability. },
    the demand is still not met,
    the typical feeder-level group of non-Aashiyana
    homes are completely cutoff from the grid.
Only if the demand is still
    not met,
    the Aashiyana Homes
    run the same algorithm.
To ensure utility maximization,
    we make homes already in a back-off state
    to use their original (iteration 1) stress level,
    where they reduce consumption to the next power-level,
    while all other homes that did not back-off
    earlier use a much lower $sl$ value for their
    threshold.This, ping-pong,
    demand reduction continues until
    demand meets supply.

\subsubsection{Centralized DLC Algorithm}

\begin{algorithm} [tp]
 \caption{Centralized DLC algorithm}
 \label{alg:CDA}
\begin{algorithmic}[1]
 \State $G_i \gets \mathbb G \> (Round Robin Selection)$
 \State$(\forall\> h_i\> \in\> \mathbb G_i\> AND\> \forall\> h_i\> \in\> \mathbb H_{na}) $ 
 \State$cl \gets L1$ 
 \State$SP = \sum h_i . cl$ 
 \State $\Delta gap \gets \Delta gap - SP$
\If{$(\Delta gap\> > \> 0)$}
   \For{$(\forall\> h_i\> \in\> \mathbb G_i\> AND\> \forall\> h_i\> \in\> \mathbb H_{a})$}
   \State$SP \gets SetToLowerLevel(h_i)$
   \State$\Delta gap \gets \Delta gap - SP$
       \If{$(\Delta gap\> > \> 0)$}
       \State $goto \> step \> 8 $
       \EndIf
   \EndFor
\EndIf
\If{$(\Delta gap\> > \> 0)$}
 \State $goto \> step \> 1 $
\EndIf
\end{algorithmic}
\end{algorithm}

Our centralized algorithm (Algorithm~\ref{alg:CDA})
    assumes that the utilities
    have a full view of the current consumption
    of each home.
The utilities (DISCOs in Pakistan)
    already have feeder-level
    groups established in which they cyclically
    implement their current load-shedding.
In our case,
    they now refine this process
    by picking the first group and computing
    the savings by shutting-off
     all non-Aashiyana homes.
With the demand-supply gap ($D-S=\Delta gap$)
    still positive,
    the DISCO starts computing energy savings
    by reducing consumption level of Aashiyana homes,
    by considering highest consumption homes first,
     to any level (chosen equally-likely)
     below their current consumption\footnote{Thus if consumption of a home is above 75\% of its meter rating, either of L4, L3, or L2 can selected.}.
They choose another home only if the $\Delta gap > 0$,
    and if cycling through all Aashiyana home
    still doesn't satisfy demand,
    the next group is selected
    and the process continues until demand is met.
Once this decision is made centrally,
    the control decisions are then communicated
    directly (and at once) to every home: Aashiyana homes are
    communicated the selected level that bypasses
    the DLC logic shown in Figure~\ref{fig:Aashiyana},
    while other homes are completely shut-off by
    their smart-meters.

\section{Evaluation of large-scale DLC using Aashiyana}
\label{sec:Eval_Results}

We now evaluate the benefits of
    our proposed large-scale DLC algorithms.
After explaining the evaluation setup,
    we compare the performance
    improvement in utility compared
    to a baseline case using
    the current schedule based load-shedding.

\subsection{Evaluation setup}
In order to best evaluate across
    a comparative behavior,
    we use an evaluation setup employing
    a custom event-driven simulator
    implemented in C++.

\subsubsection{Modeling consumption of homes}

We model the individual appliance level data
    as a distribution by using the appliance
    consumption information from the
   UK-DALE~\cite{UK-DALE_2014}
   and REDD~\cite{Redd_2011}
   dataset.
We remove the  outliers
   in the UK-DALE power consumption data-set
   by  removing values greater than
   three stdev.
We apply the Kernel Density Estimation (KDE)
   technique on this power
   consumption data  to obtain the cumulative distribution
   function (CDF) estimate, modeling a devices
   consumption in a stochastic manner.
We  then use
   inverse transform sampling to
   obtain a stochastic power consumption value
   for every device, averaged over a full hour.

We have three class of homes A, B and C
    with 7, 10, and 13 set of
    appliances respectively.
We implement the five power levels
  in the simulator as a function
  of different maximum rating for each
  class --- class A has 500W, class B has 750W,
  and class C has 1000W.
Power level 1 is full shutoff (no appliance)
   while power level 5 allows all appliances
   of that home type to
   be turned on.
We setup disconnectivity matrix for each class
    of home for every intermediate level to attain
    75\%, 50\%, and 25\% consumption of their maximum
    rating.

\begin{figure}[t]
         \centering
         \includegraphics[width=0.45\textwidth]{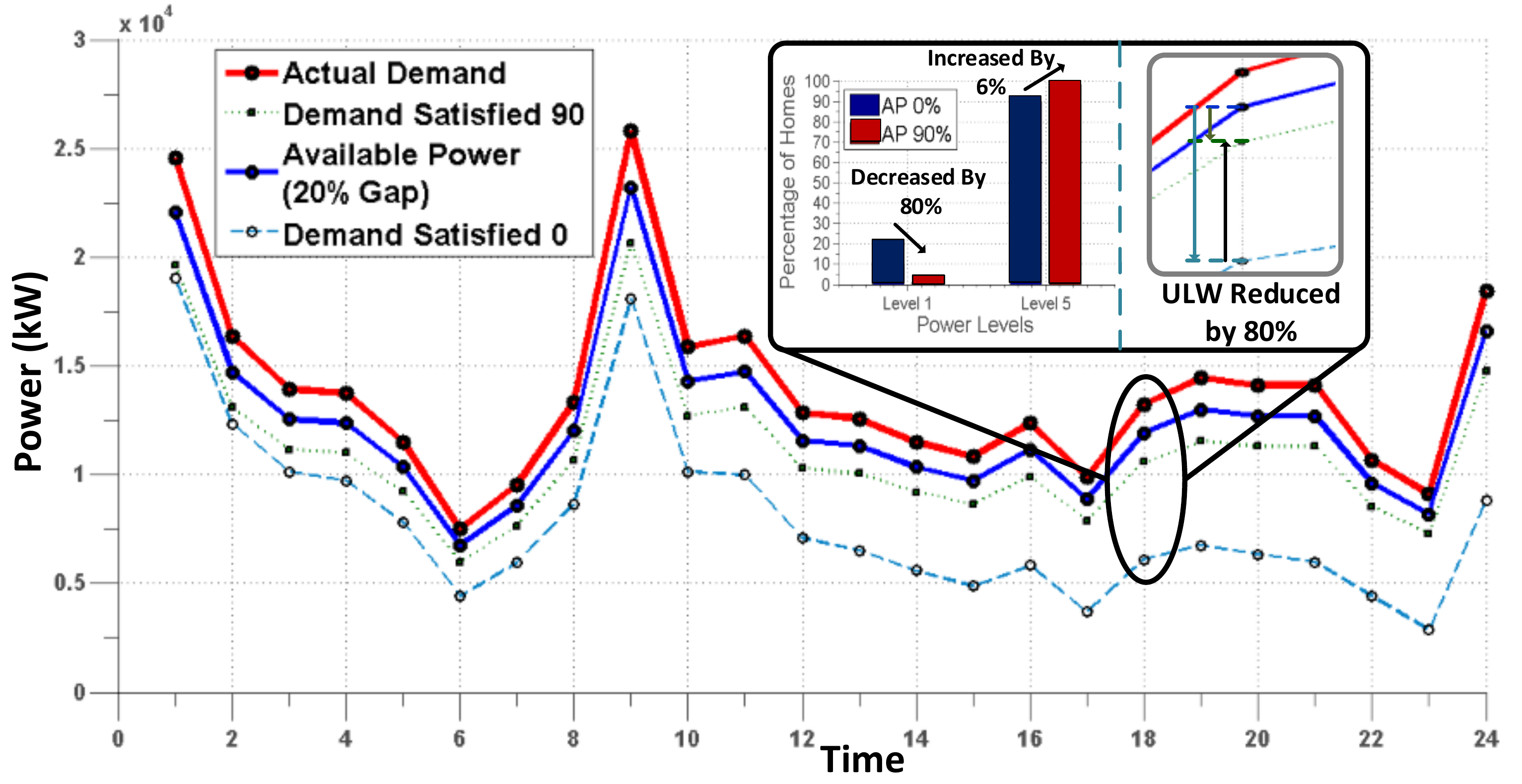}
         \caption{Observing under-load wastage (ULW) reduced by using Aashiyana-based DLC algorithms}
         \label{fig:baseline}
\end{figure}
\subsubsection{Simulator overview}
 Our custom event-driven agent-based
    simulator that operates at the granularity of
    a second.
The agents in our simulators include generators,
    distribution utilities, a central body (like NPCC)
    and home agents to simulate the electrical
    grid.
In the simulator we implemented the
    connectivity hierarchy of the DISCO
    to consumers in the grid.
We have different  Grid stations,
    feeders, transformers, and homes forming
    a tree hierarchy in the simulator.
The power consumption of different appliances
    in a home vary every hour based on their observed CDF.
Central body  checks the
    supply-demand gap every second and
    generates an event to reduce
    the gap based on the selected algorithm.
On reception of the DLC signal, homes having
    Aashiyana system will be restricted
    to limited set of appliances according to their
    respective assigned power level while
    non-Aashiyana home will be shut-off completely.
The signal of centralized DLC and turning
    of non-Aashiyana is applied to a group of ten
    feeders in the grid.
We allows these setting
     to be input as a configuration file to the
    simulator to  set the number of home agents,
    CDF data path,
    load-reduction policy,
    Aashiyana penetration (AP),
    total number
    feeders groups, and the  generation capacity.
We plan to make the code for our
    simulator open source as we believe
    it will be a great resource for the community
    to build and run large-scale simulation studies
    for smart-grid power systems.

\subsubsection{Assumptions}

We assume
    lossless transmission and distribution
    in the grid.
We also assume the utility have a
    list of all homes, whether they have
    Aashiyana installed or not.
We assumed there is no latency in reception of
    DLC signal and and that its response is
    implemented within  a second.

\subsection{Traditional Load-shedding and Under-load Wastage}

We first show the running of our simulator
    with traditional load-shedding strategy
    (selecting group of feeds to blackout, or at L1).
Figure~\ref{fig:baseline}
    shows the scenario where we
    have a 20\% demand supply gap.
We can see that there are several locations
    where the demand satisfied
    by implementing cyclic load-shedding,
    leads to something below the available generation capacity.
In these scenarios,
    the governor control on generator side
    leads to \emph{less power being generated
    than the capacity online}.
This is what we call
    under-load wastage (ULW),
    and it not only produces energy less efficiently
    across our ensemble of generators,
    but allows fewer people to have power than is possible.
While we evaluate the algorithm in detail next,
    the Figure~\ref{fig:baseline} (inset) shows
    that (for hour 18)
    our algorithms can be efficient by reducing
    ULW from 1.12MW to just 0.2MW.
This reclaimed energy allows not only a
    decrease in the number
    of home with no power (L1) by 80\% but also
     \emph{an increase} by 6\% in the fraction of homes
     with full power (L5).

\begin{figure*}[ht]
    \centering
    \begin{subfigure}[b]{0.2\textwidth}
        \centering
        \includegraphics[width=\textwidth]{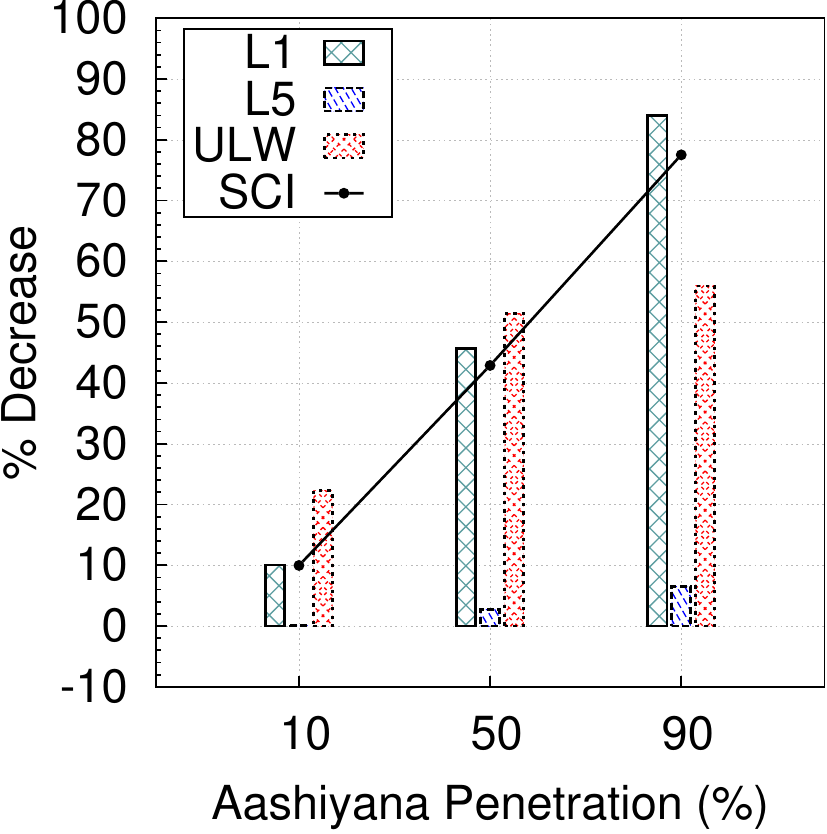}
        \caption{10\% gap}
        \label{fig:dsg10}
    \end{subfigure}
    \hfill
    \begin{subfigure}[b]{0.2\textwidth}
        \centering
        \includegraphics[width=\textwidth]{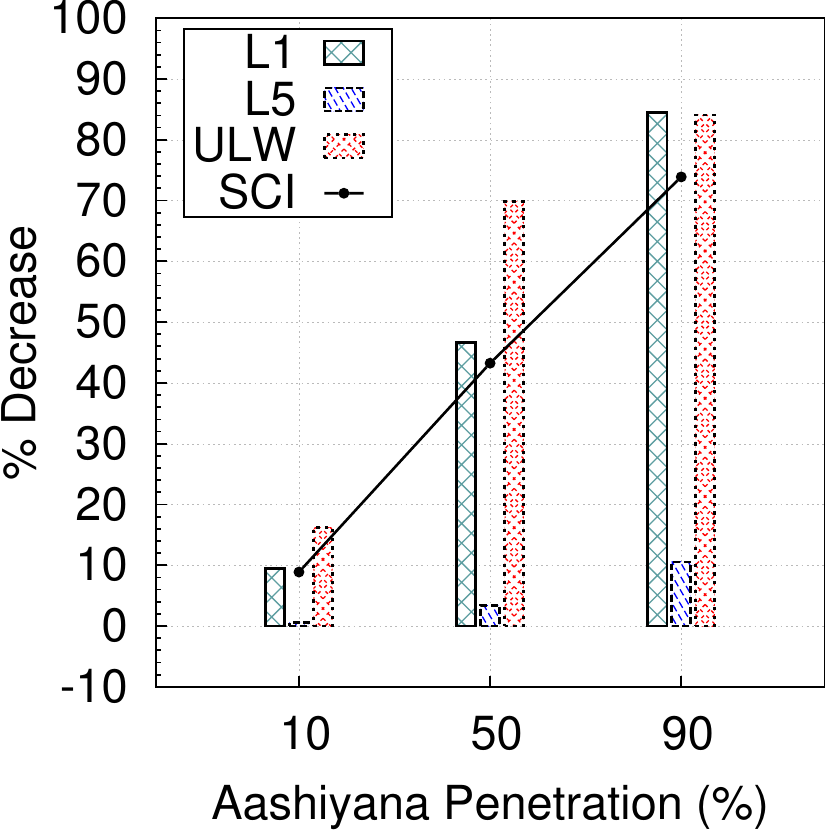}
        \caption{ 20\% gap}
        \label{fig:dsg20}
    \end{subfigure}
    \hfill
    \begin{subfigure}[b]{0.2\textwidth}
        \centering
        \includegraphics[width=\textwidth]{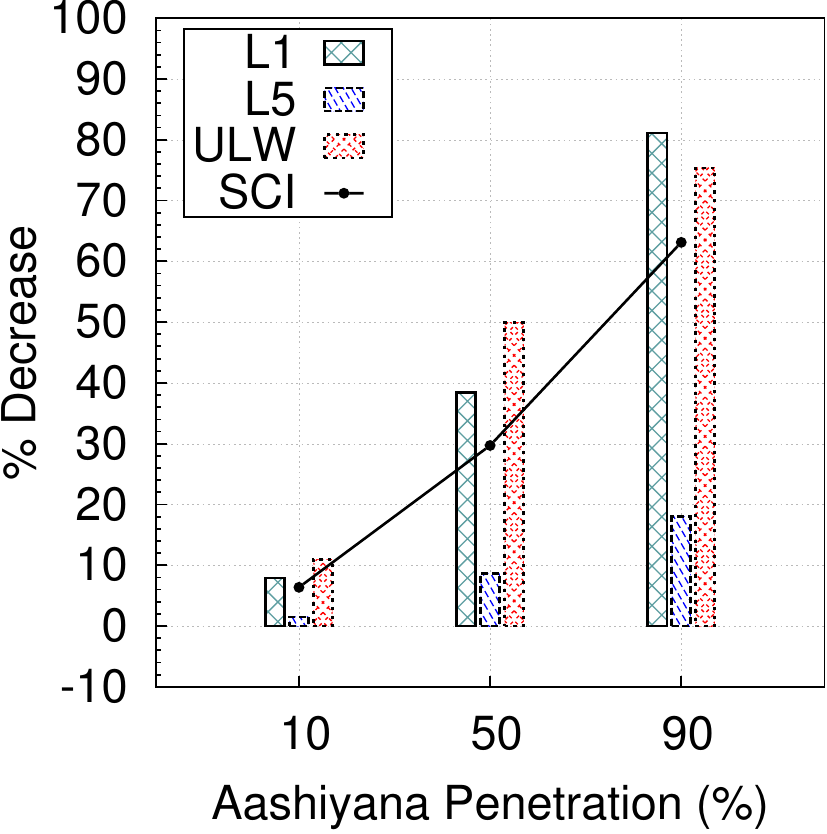}
        \caption{ 30\% gap }
        \label{fig:dsg30}
    \end{subfigure}
    \hfill
    \begin{subfigure}[b]{0.2\textwidth}
        \centering
        \includegraphics[width=\textwidth]{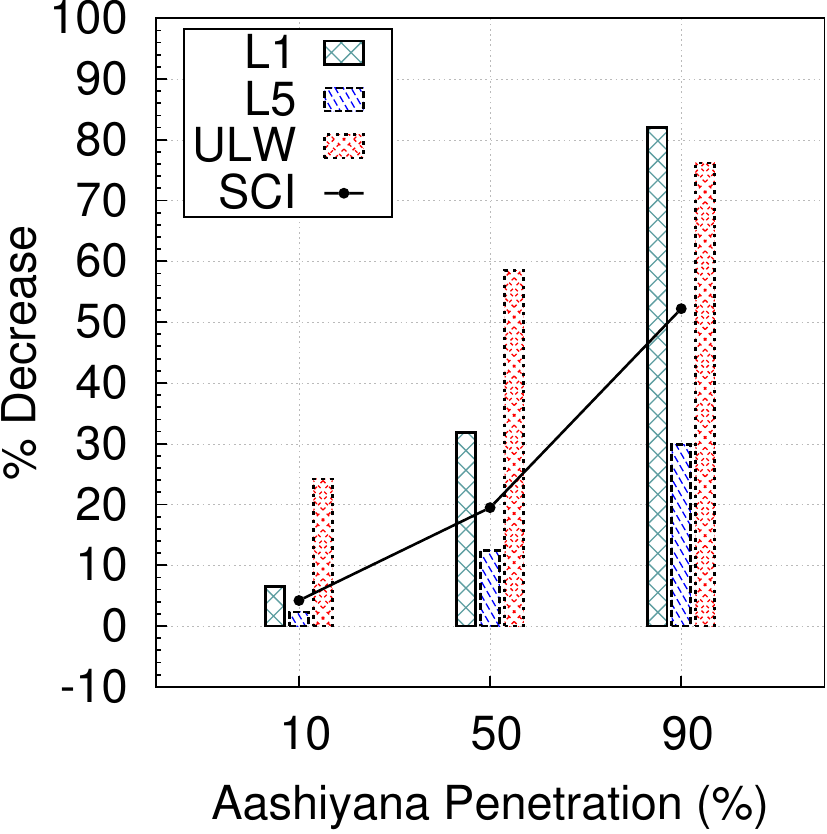}
        \caption{ 40\% gap }
        \label{fig:dsg40}
    \end{subfigure}
    \caption{Distributed Algorithm:  Change (from no AP) in distribution of home-levels with varying demand-supply gaps.}
    \label{fig:distributed_algo_results}
\end{figure*}

\subsection{Distributed Algorithm}
\label{sec:distrib_results}
Having established that for a static supply case,
    our distributed algorithm allows for increased
    social comfort by significantly reducing the number
    of homes with no-power (L1) for a slight decrease
    in homes with full-power (L5), we now
    consider a case where the supply demand gap
    is a consistent percentage of actual demand.
For this purpose,
    we vary this gap as 10-40\%,
    and observe the percentage decrease,
    over a 24hr period and compared to no
     Aashiyana penetration (AP) (i.e. traditional load-shedding),
    in homes that are in L1 and the concomitant
    decrease for L5 at different AP
    levels.

Figure~\ref{fig:distributed_algo_results}
    show the result of averaging 10 simulation runs
    for each experimental setting (\% gap, AP level).
As is quite evident,
    the fractional decrease of homes  in L1
    \emph{always} greater than (by more than 100\%)
    the corresponding decrease in L5.
This difference,
    corresponding to increase in social comfort,
    is understandably greatest at the highest
    AP level with the \textbf{social comfort index}
    (SCI\footnote{SCI is defined as the magnitude of difference between the fractional decrease in L1 and fractional decrease in L5.})
     $\approx 80$ percentage points for 90\% AP,
    thus clearly indicating the benefit
    of wide-scale adoption.
We notice that as
    the demand-gap increases,
    the improvement in SCI
    decreases.
This is so since large gap will necessarily
    demand a lots of homes to be load-shed,
    thus necessitating more homes to go below L5.

\begin{figure*}[ht]
    \centering
    \begin{subfigure}[b]{0.2\textwidth}
        \centering
        \includegraphics[width=\textwidth]{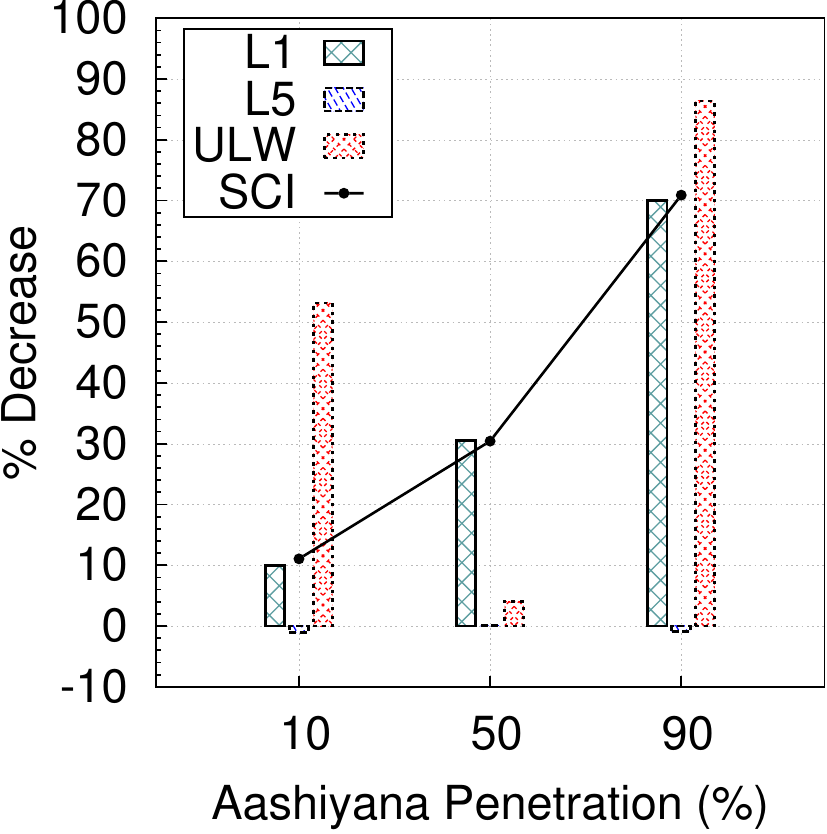}
        \caption{10\% gap}
        \label{fig:csg10}
    \end{subfigure}
    \hfill
    \begin{subfigure}[b]{0.2\textwidth}
        \centering
        \includegraphics[width=\textwidth]{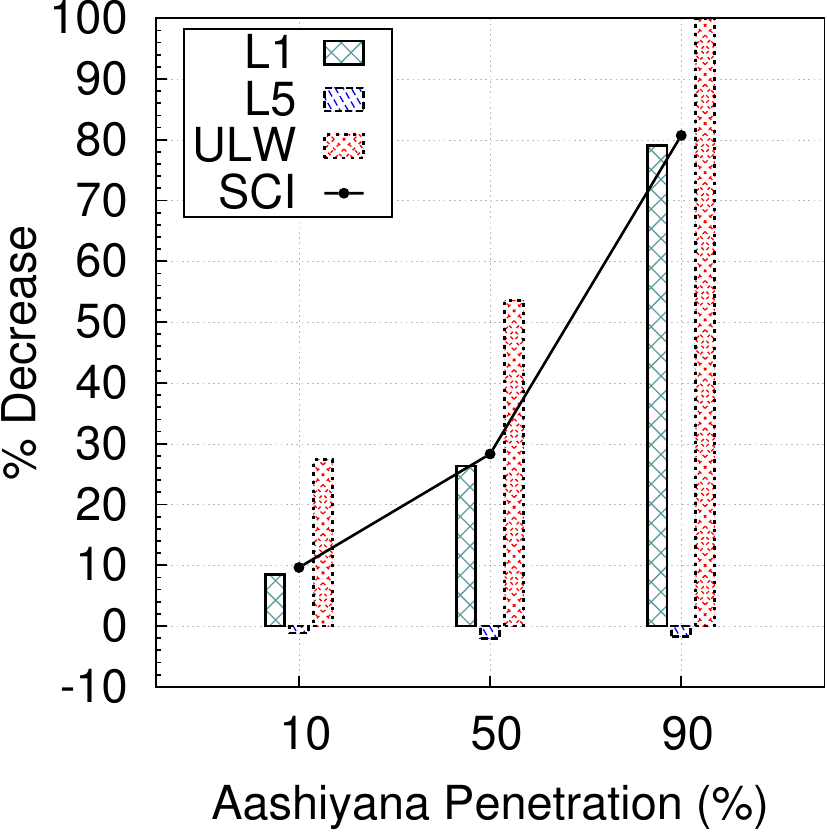}
        \caption{ 20\% gap}
        \label{fig:csg20}
    \end{subfigure}
    \hfill
    \begin{subfigure}[b]{0.2\textwidth}
        \centering
        \includegraphics[width=\textwidth]{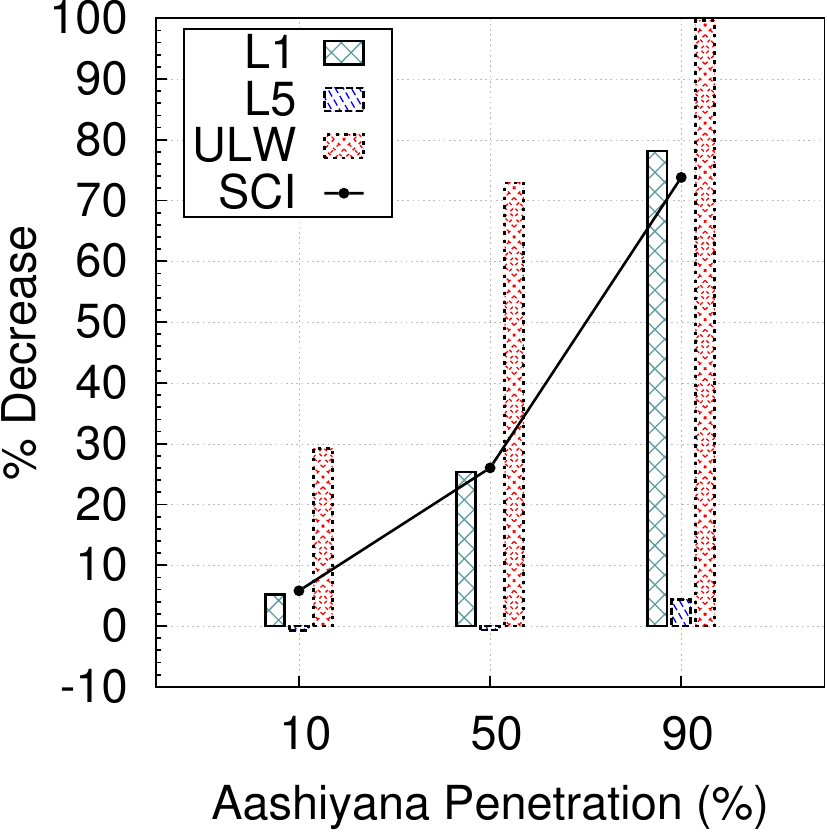}
        \caption{ 30\% gap }
        \label{fig:csg30}
    \end{subfigure}
    \hfill
    \begin{subfigure}[b]{0.2\textwidth}
        \centering
        \includegraphics[width=\textwidth]{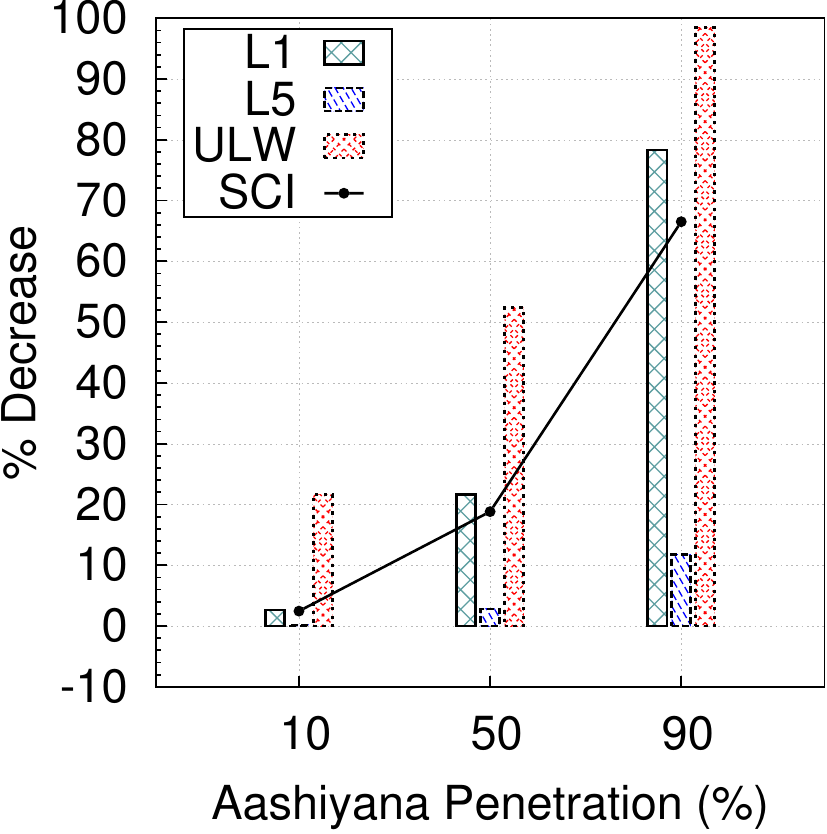}
        \caption{ 40\% gap }
        \label{fig:csg40}
    \end{subfigure}
    \caption{Centralized Algorithm: Change (from no AP) in distribution of home-levels with varying demand-supply gaps.}
    \label{fig:central_algo_results}
\end{figure*}

\subsection{Centralized algorithm}
\label{sec:central_results}

Our centralized algorithm has a holistic
    view of the energy consumption status.
We therefore expect that,
    while the trends will be similar to those
    for our distributed algorithm,
    it will be more efficient by
    reducing the amount of ULW and should
    thus help increase (SCI).

Figure~\ref{fig:central_algo_results}
    shows the results for our centralized
    algorithm.
We observe that, as before,
    most improvements are when we have
    the highest Aashiyana penetration,
    and that it decreases with increasing gap.
Further, we can confirm that
    with a more informed  strategy,
    the centralized approach obtains greater
    ULW savings, thus we have
    99\% less wastage (0.019 MW vs 3.3 MW) for 90\% AP at 20\% gap.
Moreover, it is interesting to observe that
    in certain cases, the fraction
    of homes in L5 actually increase (negative decrease).
Such a scenario is purely win-win
    since now, for the \emph{same supply-demand gap},
    we have not only decrease the fraction
    of homes that get no-power, but were also able
    to \emph{increase} the  fraction of homes
    with unrestricted power!

\subsection{Discussion}

\begin{figure}[ht]
         \centering
         \begin{subfigure}[b]{0.2\textwidth}
             \centering
             \includegraphics[width=\textwidth]{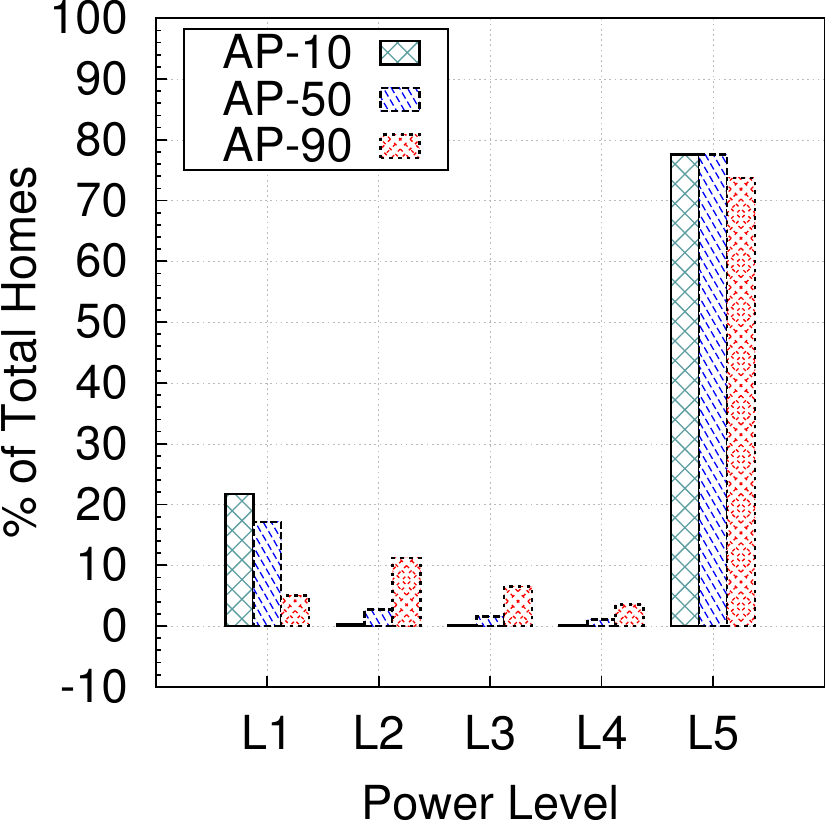}
             \caption{Centralized Algorithm}
             \label{fig:PowerLevelCentral}
         \end{subfigure}
         \hfill
         \begin{subfigure}[b]{0.2\textwidth}
             \centering
             \includegraphics[width=\textwidth]{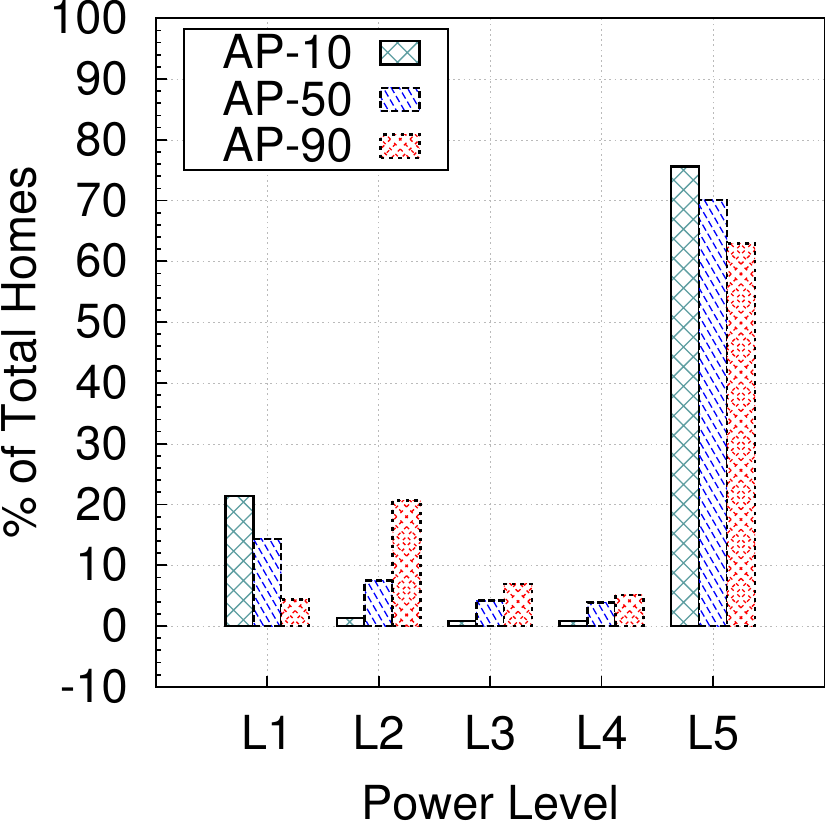}
             \caption{Distributed Algorithm}
             \label{fig:PowerLevelDist}
         \end{subfigure}
         \caption{\% of homes in all levels for a 30\% Gap.}
         \label{fig:comparison_algos}
     \end{figure}

We note that between are two algorithms,
    the centralized approach is consistently
    better at reducing the under-load wastage,
    and provides better SCI at higher AP.
However, the SCI values for the distributed case are
    slightly better at lower penetration.
We believe this is so because SCI
    show distribution for only two possible levels.
Figure~\ref{fig:comparison_algos}
    shows this distribution for homes for the same gap,
    but after different algorithms have run.
We observe that the distributed algorithm
    results in distributing savings to lift nodes
    from no-power (L1) into a low-power state.
The centralized algorithm instead,
    results in more homes with full-power
    and fewer are taken out from L1.
Using nearly any sensible definition of utility
    we see that  the centralized approach always results
    in marginally better utility for the population (around 6\% for $U(1,0.6,0.4)$).

We note however that the distributed algorithm
    can still stay close to this utility value thus
    effectively managing grid-stress.
To our mind,
    since distributed mechanisms are robust to
    individual failures, and their low complexity
    allows quick implementation,
    exploring a distributed approach to manage stress
    is a promising direction.
It is also worth exploring if increasing
    the number of levels is beneficial;
    we do caution that more levels
    decrease the user-friendliness of the system
    since the user has think about and configure
    a disconnectivity matric for each state.
%

\section{Conclusions}

We present here a novel and practical
    DLC system, Aashiyana, that enables
    several different low-power states
    for homes within the context of
    highly stressed grids.
We design and implement
    this with practical
    incentives for the utilities
    (decreasing social unrest)
    as well as consumers
    (low-cost, lower hours with no-power, greater utility),
    \emph{all without} having to increase
    the supply side equation.
We propose two types of algorithms
    that utilize this ability
    of guaranteed budget reduction at different levels
    that allow for more efficient reduction in gap
    with reduced amount of underload wastage.
We show that, compared to current load-shedding strategy,
    for the same supply-demand gap,
    we can reduce
    homes with \emph{no} power by $>80$\%
    while not significantly impacting
    the fraction of homes with full power.

\paragraph{Acknowledgements}

We want to thank Dr. Mukhtar Ullah for helping design
    the stochastic model of appliance consumption and Zaafar Ahmed for supporting in generating graphs.

\bibliographystyle{abbrv}
\bibliography{paper}

\end{document}